\documentclass[11pt]{article}
\usepackage{amssymb, latexsym, amsmath, hyperref}
\usepackage{epsfig}
\usepackage{graphicx}
\usepackage{amsfonts}
\usepackage{mathrsfs}
\usepackage{multirow}
\usepackage[dvips]{color}

\newcommand{\R}{\mathbb{R}}

\newcommand{\fn}{{\,\mathfrak{n}\,}}
\newcommand{\ft}{\mathfrak{t}}

\newcommand{\fz}{\mathfrak{z}}

\newcommand{\cA}{{\mathcal{A}}}

\newcommand{\cF}{\mathcal{F}}

\newcommand{\be}{\begin{equation}}
\newcommand{\ee}{\end{equation}}
\newcommand{\bea}{\begin{eqnarray}}
\newcommand{\eea}{\end{eqnarray}}
\newcommand{\nn}{\nonumber}

\newcommand{\ed}{\end{document}}

\newcommand{\rz}{{\mbox{\scriptsize${\rm Z}$}}}

\newcommand{\bi}{\begin{itemize}}
\newcommand{\ei}{\end{itemize}}

\newcommand{\bce}{\begin{center}}
\newcommand{\ece}{\end{center}}

\newcommand{\RE}{{\rm Re}}
\newcommand{\IM}{{\rm Im}}

\newcommand{\od}{\overleftrightarrow{\partial}}

\begin{document}

\title{Optical Spectral Singularities and Coherent Perfect Absorption in a Two-Layer Spherical Medium}

\author{Ali~Mostafazadeh\thanks{E-mail address:
amostafazadeh@ku.edu.tr, Phone: +90 212 338 1462, Fax: +90 212 338
1559}~~and Mustafa Sar{\i}saman\thanks{E-mail address:
msarisaman@ku.edu.tr, Phone: +90 212 338 1378, Fax: +90 212 338
1559}
\\
Department of Mathematics, Ko\c{c} University, \\ 34450 Sar{\i}yer,
Istanbul, Turkey}

\date{ }
\maketitle

\begin{abstract}

An optical spectral singularity is a zero-width resonance that corresponds to lasing at threshold gain. Its time-reversal causes coherent perfect absorption of light and forms the theoretical basis of antilasing. In this article we explore optical spectral singularities of a two-layer spherical medium. In particular, we examine the cases that a gain medium is coated by a thin layer of high-refractive index glass and a spherical glass covered by a layer of gain material. In the former case, the coating reduces the minimum radius required for exciting spectral singularities and gives rise to the formation of clusters of spectral singularities separated by wide spectral gaps. In the latter case, the coating leads to a doubling of the number of spectral singularities.

\vspace{2mm}

\vspace{2mm}

\end{abstract}

\section{Introduction}

The discovery \cite{prl-2009} that the mathematical concept of a spectral singularity \cite{ss} has physical realizations as zero-width resonances of complex scattering potentials has motivated a detailed study of this phenomenon \cite{pra-2009} -- \cite{samsonov}. In particular, it is shown that optical spectral singularities (OSS) correspond to the lasing at the threshold gain \cite{pra-2011a} and that a time-reversed optical spectral singularity \cite{longhi-phys-10} yields a coherent perfect absorption (CPA) of light, i.e., an antilaser \cite{antilaser}.

Typical lasers are photonic devices consisting of an active medium
placed inside an optical cavity. A particularly interesting type of
lasers are those based on spherical granules where the surface of
the sphere acts as the cavity \cite{spherical-lasers}. These are
characterized by their extremely high quality factors and small
volumes of excitation. Owing to these properties, a spherical laser
can also be realized in which the active medium is located outside
an spherical core \cite{Matsko}.

\begin{figure}
      \begin{center}
      \includegraphics[scale=.35]{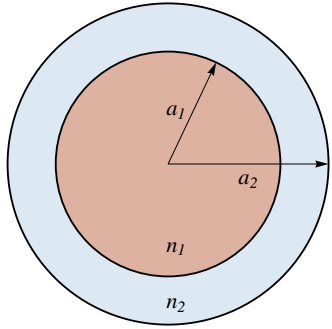}\\
      \caption{(Color online) Two-layer spherical gain medium. $a_1$ and $a_2$ are the inner and outer radii of the spherical shell; $\fn_{1}$ and $\fn_{2}$ are the complex refractive indices of the inner core and the author shell, respectively.}
      \label{figure1}
      \end{center}
      \end{figure}

In Ref.~\cite{pla-2011} we studied the OSS of a uniform spherical
gain medium and showed for the radial (transverse) modes of a
concrete spherical dye laser that the emergence of an OSS puts a
lower bound on the radius of the gain medium. This is the minimum
radius $a^{(\rm min)}$ required for lasing in these modes. In the
present article we examine OSS and CPA for an active medium
consisting of a spherical inner core and a spherical outer shell
with different refractive indices, as shown in Figure~\ref{figure1}.
In particular, we wish to explore the prospects of reducing the
value of $a^{(\rm min)}$ by means of coating the spherical gain
medium by a material with higher refractive index.\footnote{It is
well-known that the presence of such a coating increases the
system's quality factor \cite{Matsko,Sandber}.}

\section{Radial Transverse Spherical Electromagnetic Waves}

Consider an optically active material with an inner spherical core of radius $a_1$ and an outer spherical shell of thickness $a_2-a_1$ placed in vacuum. Let $\fn_1$ and $\fn_2$ denote the complex refractive indices of the inner core and the outer shell, respectively, and suppose that they are independent of space and time. The electromagnetic (EM) waves interacting with this system satisfy the Maxwell equations:
    \begin{equation}
    \vec{\nabla}\cdot\vec{D} = 0, \ \ \ \ \ \ \ \ \ \ \ \ \ \ \ \ \
    \vec{\nabla}\cdot\vec{B} = 0, \label{equation1}
    \end{equation}
    \begin{equation}
    \vec{\nabla} \times \vec{E} + \dot{\vec{B}} = 0, \ \ \ \ \ \ \ \ \ \
    \ \vec{\nabla} \times \vec{H} - \dot{\vec{D}} = 0,
    \label{equation2}
    \end{equation}
where $\vec{D} := \varepsilon_0\fz(r) \vec{E}$,
$\vec{H} := \mu_0^{-1} \vec{B}$,  $\varepsilon_0$ and $\mu_0$ are respectively the permeability and permittivity of the vacuum, $r:=|\vec r|$ is the radial spherical coordinate,
    \be
    \fz(r):=\left\{\begin{array}{ccc}
    \fn_{1}^2 & {\rm for} & r<a_1,\\
    \fn_{2}^2 & {\rm for} & a_1\leq r<a_2,\\
    1 & {\rm for} & r\geq a_2,
    \end{array}\right.
    \label{e1}
    \ee
and each over-dot represents a time-derivative. According to (\ref{equation2}), the electric field $\vec E=\vec{E}(\vec{r}, t)$ is a solution of the wave equation:
    \begin{equation}
    \ddot{\vec{E}}(\vec r,t) + \Omega^{2} \vec{E}(\vec r,t) = 0,~~~~~~r\neq a_1,a_2,
    \label{equation3}
    \end{equation}
where $\Omega^{2} :=  c^2\fz(\vec{r})^{-1}\vec{\nabla} \times \vec{\nabla} \times$ and $c=(\varepsilon_0\mu_0)^{-1/2}$ is the speed of light in vacuum.

For a time-harmonic EM field with angular frequency $\omega$ that propagates in a charge-free medium, we have $\vec{E}(\vec{r}, t) = e^{-i\omega t}\vec{E}(\vec{r})$, and  (\ref{equation3}) reduces to the time-independent Schr\"odinger equation:
    \begin{equation}
    -\nabla^{2} \vec{E}(\vec{r}) +  v (r) \vec{E}(\vec{r})
    = k^{2} \vec{E}(\vec{r}),
    \label{equation4}
    \end{equation}
where  $k := \omega/c$ is the wave number and $v$ is the complex
barrier potential: $v(r):=k^{2}[1-\mathfrak{z}(r)]$.\footnote{Note
that (\ref{equation4}) is nothing but the well-known Helmholtz
equation, and the potential appearing in it is energy-dependent. For
the purposes of our investigation this does not cause any
difficulty.}

Following the analysis of Ref.~\cite{pla-2011}, we investigate transverse radially propagating spherical solutions of (\ref{equation4}) that have the form
    \be
    \vec{E} (\vec{r}) =E(r) \hat{\phi}.
    \label{e2}
    \ee
Here $\hat\phi$ is the unit vector associated with the azimuthal angular coordinate $\phi$ of the spherical coordinate system. Inserting (\ref{e2}) in (\ref{equation4}) yields
    \begin{equation}
    \left[\frac{d^{2}}{dr^{2}} + \frac{2}{r}\frac{d}{dr} + k^{2}
    -v(r) -\frac{1}{r^{2}}\right]E(r)= 0.
    \label{equation5}
    \end{equation}
For $r<a_1$, $a_1<r<a_2$ and $r>a_2$ where $v$ takes constant values, we can transform (\ref{equation5}) to the spherical Bessel equation of order $\nu:= \frac{\sqrt{5}}{2}$, \cite{jackson}. Therefore,
    \be
    E(r)=\left\{\begin{array}{ccc}
    A_1 j_\nu(k_1r) + B_1 n_\nu (k_1r) &{\rm for}& r<a_1,\\
    A_2 j_\nu(k_2r) + B_2 n_\nu (k_2r) &{\rm for}& a_1<r<a_2,\\
    A_3 h_\nu^{(1)} (kr) + B_3 h_\nu^{(2)}(kr)&{\rm for}&  r>a_2,\end{array}\right.
    \label{e3}
    \ee
where the $A_i$ and $B_i$ are constant numerical coefficients, $j_\nu, n_\nu$ and $h_\nu^{(i)}$ are respectively the spherical Bessel, Neumann, and Hankel functions, and $k_{i} :=\fn_{\!\!i}\,k$ for $i=1,2$. Notice that the condition that the electric field be regular at the origin implies that $B_1=0$. Consequently, $E(0)=0$.

Having obtained the explicit form of the electric field, we can compute the magnetic field using (\ref{equation2}). This gives
    \be
    \vec{B}(\vec r,t)=e^{-i\omega t}B(r)\hat\theta,
    \label{e4}
    \ee
where $\hat\theta$ is the unit vector associated with  the spherical polar coordinate $\theta$,
    \be
    B(r):=i\,\omega^{-1}\tilde E(r)=\left\{\begin{array}{ccc}
    i\,\omega^{-1}
    \left[A_1 \tilde j_\nu (k_1r)+ B_1 \tilde n_\nu(k_1r)\right] &{\rm for}& r<a_1,\\
     i\,\omega^{-1}
    \left[A_2 \tilde j_\nu (k_2r)+ B_2 \tilde n_\nu(k_2r)\right] &{\rm for}& a_1<r<a_2,\\
     i\,\omega^{-1}
    \left[A_3 \tilde h_\nu^{(1)} (kr) + B_3 \tilde h_\nu^{(2)}(kr)\right]&{\rm for}&  r>a_2,\end{array}\right.
    \label{e5}
    \ee
and for each differentiable function  $f$ we define $\tilde f$ according to
$\tilde f(r):=\left(\frac{d}{dr}+\frac{1}{r}\right)f(r)$. Notice than for every pair of differentiable functions $f,g:\R^+\to\R$ and positive real numbers $r,s\in\R^+$,
    \be
    f(r)\tilde g(s)-g(s)\tilde f(r)=f(r)g'(s)-g(s)f'(r).
    \label{identity1}
    \ee
Because this quantity will frequently appear in our calculations, we denote it by $f(r)\overleftrightarrow{\partial}g(s)$ for brevity. In other words,
    \be
    f(r)\overleftrightarrow{\partial}g(s):=f(r)g'(s)-g(s)f'(r)=
    f(r)\tilde g(s)-g(s)\tilde f(r).
    \label{notation1}
    \ee
Recall that the Wronskian of $f$ and $g$ is given by $W[f(r),g(r)]:=f(r)\od g(r)$. Therefore
    \be
    f(r)\tilde g(r)-g(r)\tilde f(r)=W[f(r),g(r)].
    \label{wronskian}
    \ee

In order to relate the coefficients $A_j$ and $B_j$ appearing in (\ref{e3}), we need to impose the appropriate matching conditions at the boundaries $r=a_1$ and $r=a_2$, \cite{jackson}. For the system we consider, these correspond to the condition that the parallel component of both the electric and magnetic fields must be continuous at the boundaries. In view of (\ref{e2}) and (\ref{e4}), this means that $E=E(r)$ and $B=B(r)$ must be continuous functions. We can satisfy this condition provided that we select the coefficients $A_i$ and $B_i$ such that $E$ and $B$ are continuous at $r=a_1$ and $r=a_2$. This gives
    \be
   \mathbf{K}_{11}\left[\begin{array}{c}
    A_{1} \\
    0\end{array}\right]= \mathbf{K}_{12}\left[\begin{array}{c}
    A_{2} \\
    B_{2}\end{array}\right],~~~~~~
   \mathbf{K}_{22} \left[\begin{array}{c}
    A_{2} \\
    B_{2} \end{array}\right]=
    \mathbf{L} \left[\begin{array}{c}
    A_3 \\
    B_3 \end{array}\right],
    \label{e7}
    \ee
where for all $p,q=1,2$,
    \be
    \mathbf{K}_{pq}:= \left[
    \begin{array}{cc}
    j_{\nu} (k_{q}a_p) & n_{\nu} (k_{q}a_p) \\
    \tilde j_{\nu} (k_{q}a_p) & \tilde n_{\nu} (k_{q}a_p)
    \end{array}\right],~~~~~~
    \mathbf{L}:= \left[
    \begin{array}{cc}
    h_{\nu}^{(1)} (ka_2) & h_{\nu}^{(2)} (ka_2) \\
    \tilde h_{\nu}^{(1)} (ka_2) & \tilde h_{\nu}^{(2)} (ka_2)
    \end{array}\right].
    \label{KL}
    \ee
According to these equations and (\ref{wronskian}),
    \bea
    \det(\mathbf{K}_{pq})&=&
    W[j_\nu(x),n_\nu(x)]\Big|_{x=k_q a_p}=(k_q a_p)^{-2}\neq 0,
    \label{W1}\\
    \det(\mathbf{L})&=&
    W[h^{(1)}_\nu(x),h^{(2)}_\nu(x)]\Big|_{x=ka_2}=-i(ka_2)^{-2}\neq 0.
    \label{W2}
    \eea
The fact that these quantities do not vanish was to be expected,
because $(j_\nu,n_\nu)$ and $(h^{(1)},h^{(2)})$ are pairs of
linearly-independent solutions of a second order homogeneous linear
differential equation. Eqs.~(\ref{W1}) and (\ref{W2}) imply that
$\mathbf{K}_{pq}$ and $\mathbf{L}$ are invertible matrices. We can
use their inverse together with Eq.~(\ref{e7}) and the fact that
$A_1\neq 0$, to express $A_3$ and $B_3$ in terms of $A_1$ according
to $A_3=M_{11} A_1$ and $B_3=M_{21} A_1$, where $M_{ij}$ are the entries of the transfer matrix
    \be
    \mathbf{M}:=\mathbf{L}^{-1}\mathbf{K}_{22}
    \mathbf{K}_{12}^{-1}\mathbf{K}_{11}.
    \label{e9}
    \ee

\section{OSS and CPA for Radial Transverse Spherical Waves}

As discussed in Ref.~\cite{pla-2011}, we can easily exploit the asymptotic properties of the spherical Hankel functions to infer that the reflection amplitude of our system is given by
    \be
    R:=\frac{A_3}{B_3}=\frac{M_{11}}{M_{21}}.
    \label{e10}
    \ee
Therefore, in order for characterizing OSS and CPA that correspond to the real poles and zeros of $R$, we only need to calculate $M_{11}$ and $M_{21}$. Using (\ref{identity1}), (\ref{W1}), (\ref{W2}), (\ref{notation1}), (\ref{KL}), (\ref{e9}),
and doing the necessary algebra, we obtain
    \be
    M_{11}= N_2,~~~~M_{21}=-N_1,
    \label{M11}
    \ee
where for both $\ell=1,2$,
    \bea
    &&N_\ell:=
    P\left[n_\nu(k_2a_2)\od h^{(\ell)}_\nu(ka_2)\right]+
    Q\left[ j_\nu(k_2a_2)\od h^{(\ell)}_\nu(ka_2)\right],
    \label{N=}\\
    &&P:=i(kk_2a_1a_2)^2 j_\nu(k_2a_1)\od j_\nu(k_1a_1),~~~~
    Q:=i(kk_2a_1a_2)^2 j_\nu(k_1a_1)\od n_\nu(k_2a_1).
    \label{PQ}
    \eea
In view of  (\ref{M11}) -- (\ref{PQ}) and the fact that $f\od g=-g\od f$, we can express (\ref{e10}) in the form
    \be
    R = -
    \frac{\big[j_\nu(k_2a_1)\od j_\nu(k_1a_1)\big]
    \big[n_\nu(k_2a_2)\od h^{(2)}_\nu(ka_2)\big]-
    \big[n_\nu(k_2a_1)\od j_\nu(k_1a_1)\big]
    \big[j_\nu(k_2a_2)\od  h^{(2)}_\nu(ka_2)\big]}{
    \big[j_\nu(k_2a_1)\od j_\nu(k_1a_1)\big]
    \big[n_\nu(k_2a_2)\od h^{(1)}_\nu(ka_2)\big]-
    \big[n_\nu(k_2a_1)\od j_\nu(k_1a_1)\big]
    \big[j_\nu(k_2a_2)\od  h^{(1)}_\nu(ka_2)\big]}.
    \nn
    \ee

Eqs.~(\ref{M11}) provide a simple demonstration of the fact that CPA corresponds to an OSS of the time-reversed system. To see this we recall that we can obtain the time-reversed system by complex-conjugating the refractive indices $\fn_\ell$. This implies
    \bea
    &k_\ell\to k_\ell^*,~~~~P\to P^*,~~~~Q\to Q^*,~~~~N_1\to N_2^*,~~~~
    N_2\to N_1^*,&\\
    &M_{11}\to -M_{21}^*,~~~~~~M_{21}\to -M_{11}^*,~~~~~~ R\to \mbox{\large$\frac{1}{R^{*}}$},&
    \label{tt}
    \eea
where we have used (\ref{N=}), (\ref{PQ}), (\ref{M11}), (\ref{e10}), and the fact that $h_\nu^{(1)}(ka_2)^*=h_\nu^{(2)}(ka_2)$.  According to the last relation in (\ref{tt}), a CPA, that corresponds to $R=0$, appears if and only if the reflection coefficient of the time-reversed system diverges, i.e., the latter develops an OSS and begins lasing at the threshold gain . Therefore, in the following we only consider the problem of locating OSS. We can easily obtain the values of the physical parameters leading to a CPA of the spherical waves we consider by complex-conjugating the refractive indices $\fn_1$ and $\fn_2$ or changing the sign of the gain/attenuation coefficients of both the layers.

A straightforward consequence of this observation is the fact that because the law of energy conservation prohibits the emergence of an OSS for the case that both the interior core and the outer shell of our system consist of lossy material,  a CPA cannot be realized unless either the core or the outer shell includes a lossy medium. Furthermore, it is possible to generate both a CPA and an OSS, if our system involves both lossy and gain media.

In order to determine the location of OSS in the space of the physical parameters of the system, we study the real zeros of $M_{21}$ or alternatively $N_1$ in the complex $k$-plane. In view of (\ref{N=}) and (\ref{PQ}), this is equivalent to finding the real values of $k$ fulfilling
    \be
    \big[j_\nu(k_2a_1)\od j_\nu(k_1a_1)\big]
    \big[n_\nu(k_2a_1)\od h^{(1)}_\nu(ka_2)\big]=
    \big[n_\nu(k_2a_1)\od j_\nu(k_1a_1)\big]
    \big[j_\nu(k_2a_1)\od h^{(1)}_\nu(ka_2)\big].
    \label{exactspect}
    \ee
Noting that $k_i=\fn_i k$, this is a complex transcendental equation involving two complex variables, namely $\fn_1$ and $\fn_2$, and two (positive) real variables: $x_1:=ka_1$ and $x_2:=ka_2$.

For $a_1=a_2$ and $\fn_1=\fn_2$ that corresponds to a homogeneous spherical medium that we consider in \cite{pla-2011}, the first factor on the left-hand side of (\ref{exactspect}) vanished identically and the first factor on its right-hand side becomes $W[n_\nu(x),j_\nu(x)]$ with $x=k_1a_1=x_1\fn_1$. Because the latter is nonzero, (\ref{exactspect}) reduces to
    \be
    j_\nu(k_1a_1)\od h^{(1)}_\nu(ka_1)=0.
    \label{single}
    \ee
We can use the identity:
    \be
    u'_\nu(\rz)= \frac{\nu \,u_{\nu-1}(\rz) - (\nu+1)u_{\nu+1}(\rz)}{2\nu + 1},
    \label{recur}
    \ee
to express the derivative of the spherical Bessel, Neumann, and Hankel functions. Doing this in (\ref{single}) gives rise to the equation for the spectral singularities of a spherical gain medium that we derive in \cite{pla-2011}, namely
    \be
    \frac{d}{dr} \ln h_{\nu}^{(1)}(kr)\Big|_{r=a_{1}} = \frac{d}{dr} \ln
    j_\nu (k_{1} r)\Big|_{r=a_{1}}.
    \label{equation18}
    \ee

Employing the identity (\ref{recur}) in (\ref{exactspect}) gives a
more lengthy equation for OSS that we use in our numerical and
graphical investigations. Before reporting the results of this
investigation, however, we will carry out a perturbative analysis of
(\ref{exactspect}). Similarly to the single-layer spherical medium
we studied in \cite{pla-2011}, this turns out to reveal some basic
properties of the solutions.

\section{Perturbative Analysis of OSS}

Consider the Mie regime where $a_2\geq a_1\gg 2\pi/k=:\lambda$ and $|\fn_i|<4$. Then $x_2\geq x_1\gg 1$,  $|k_ia_1|=|\fn_i|x_1=2\pi |\fn_i|a_1/\lambda \gg 1$, and we can perform a large-$x_1$ (and -$x_2$) expansion of the terms appearing in (\ref{exactspect}). This requires using the following asymptotic expansions of the spherical Bessel, Neumann, and Hankel functions.
    \bea
     j_\nu (\rz) &=& \frac{\sin(\rz - \frac{\pi\nu}{2})}{\rz}
     \sum_{s=0}^{\infty} \frac{(-1)^{s}\cA_{2s}(\nu)}{\rz^{2s}} + \frac{\cos(\rz-\frac{\pi\nu}{2})}{\rz} \sum_{s=0}^{\infty} \frac{(-1)^{s}\cA_{2s+1}(\nu)}{\rz^{2s+1}},
     \label{asymp-j}\\
     n_\nu (\rz) &=& -\frac{\cos(\rz - \frac{\pi\nu}{2})}{\rz}
     \sum_{s=0}^{\infty} \frac{(-1)^{s}\cA_{2s}(\nu)}{\rz^{2s}} + \frac{\sin(z-\frac{\pi\nu}{2})}{\rz} \sum_{s=0}^{\infty} \frac{(-1)^{s}\cA_{2s+1}(\nu)}{\rz^{2s+1}},
     \label{asymp-n}\\
     h_{\nu}^{(\ell)}(\rz)&=& \frac{e^{-i(-1)^\ell(\rz-\frac{\pi\nu}{2})}}{\rz}  \left[ i(-1)^\ell \sum_{s=0}^{\infty} \frac{(-)^{s}\cA_{2s}(\nu)}{\rz^{2s}}+
     \sum_{s=0}^{\infty} \frac{(-1)^{s}\cA_{2s+1}(\nu)}{\rz^{2s+1}}\right], \label{asymp-h2}
    \eea
where
    \be
    \cA_k(\nu):=\frac{\Gamma(\nu+k+1)}{2^{k} k! \Gamma(\nu-k+1)}=
    \frac{\prod_{\ell=0}^{2k-1}(\nu+k-\ell)}{2^{k} k!},\nn
    \ee
and $\Gamma$ stands for the Gamma function.

Substituting (\ref{asymp-j}) -- (\ref{asymp-h2}) in (\ref{exactspect}), neglecting the quadratic and higher order terms in $x_1^{-1}$ and $x_2^{-1}$ in the resulting equation, noting that $\cA_0(\nu)=1$ and $\cA_1(\nu)=\frac{\nu(\nu+1)}{2}$, and introducing
    \bea
    &&\ft_\ell:=\tan(k_\ell\, a_1-\frac{\pi\nu}{2}) =\tan(x_1\fn_\ell-\frac{\pi\nu}{2}),
    \label{u-v}\\
    &&
    \cF(x_1,\fn_1,\fn_2):=\frac{\fn_1(\fn_2+i\ft_2)+\fn_2\ft_1(\fn_2\ft_2-i)}{
    \fn_1(i-\fn_2\ft_2)+\fn_2\ft_1(\fn_2+i\ft_2)},
    \label{cF=}
    \eea
we find
    \be
    \tan(x_2\fn_2-\frac{\pi\nu}{2})\approx \cF(x_1,\fn_1,\fn_2),
    \label{equation17}
    \ee
where we use ``$\approx$'' to indicate that this equation is obtained by employing first-order perturbation theory. Solving (\ref{equation17}) for $x_2$ yields
    \bea
    x_2\approx\frac{\pi\nu +2\tan^{-1}(\cF)}{2\fn_2}=
    \frac{1}{2\fn_2}\left[\pi \left(2m+\nu\right)-i\ln\left(\frac{1+i\cF}{1-i\cF}\right)\right],
    \label{a2=}
    \eea
where we have used ``$\ln$'' to denote the principal part of natural logarithm of its argument, suppressed the argument of $\cF$ for brevity, and employed the identity
    \be
    \tan^{-1}(\rz)=\pi m+\frac{1}{2i}\ln\left(\frac{1+i\rz}{1-i\rz}\right),~~~~m=0,\pm1,\pm2,\cdots.
    \label{identity}
    \ee
The parameter $m$ appearing in (\ref{a2=}) is an integer that we identify with a mode number labeling OSS.

For a homogeneous spherical medium, where $a_1=a_2$ and $\fn_1=\fn_2$, we have $x_1=x_2$, $\cF=-i\fn_1$, and (\ref{a2=}) gives
    \be
    x_1\approx\frac{1}{2\fn_1}\left[
    \pi \left(2m+ \nu+1 \right)-i
    \ln\left(\frac{\fn_1+1}{\fn_1-1}\right)\right]. \label{e41}
    \ee
This is in complete agreement with Eq.~(23) of Ref.~\cite{pla-2011}.

Because the left-hand side of (\ref{a2=}) is real, we can express this complex equation as a pair of real equations:
    \bea
    &&x_2-\RE\left\{\frac{1}{2\fn_2}
    \left[\pi (2m+\nu)-i    \ln\left(\frac{1+i\cF}{1-i\cF}\right)\right]\right\}\approx0,
    \label{eq1z}\\
    &&\IM\left\{\frac{1}{\fn_2}
    \left[\pi (2m+\nu)-i \ln\left(\frac{1+i\cF}{1-i\cF}\right)\right]\right\}\approx0.
    \label{eq2z}
    \eea
We also note that
    \be
    \frac{1+i\cF}{1-i\cF}=
    \frac{(\fn_2\ft_1+i\fn_1)(1+i\ft_2)(\fn_2+1)}{
    (\fn_2\ft_1-i\fn_1)(1-i\ft_2)(\fn_2-1)}.
    \label{arg=}
    \ee

Next, we consider the special case of a coated spherical active medium with $a_2-a_1\ll a_1$. In this case the gain/absorption properties of the coating can be neglected and $\fn_2$ may be assumed to take a real value that we label by $n_2$. This implies the equivalence of $(\ref{eq2z})$ and the condition that the absolute value of (\ref{arg=}) must be unity. Imposing this condition and noting that in this case $\ft_2$ is also real, we find
    \be
    \left|\frac{n_2\ft_1+i\fn_1}{n_2\ft_1-i\fn_1}\right|\approx \frac{|n_2-1|}{n_2+1}.
    \label{eq3z}
    \ee
This is a real equation involving a complex variable, $\fn_1$, and two real variables, $n_2$ and $x_1$. In particular, it does not involve the mode number $m$.

In order to use (\ref{eq3z}) for locating OSS, we express $\fn_1$ and $\ft_1$ in terms of their real and imaginary parts. Let $\eta_1$ and $\kappa_1$ denote the real and imaginary parts of $\fn_1$, so that
    \be
    \fn_1=\eta_1+i\kappa_1,
    \label{n1=}
    \ee
and introduce
    \be
    \alpha:=x_1\eta_1- \frac{\pi\nu}{2}
    = k a_1\eta_1- \frac{\pi\nu}{2} =2\pi\left(\frac{a_1\eta_1}{\lambda}-\frac{\nu}{4}\right)
    ,~~~~~
    \beta:=x_1\kappa_1=ka_1\kappa_1=\frac{2\pi a_1\kappa_1}{\lambda}.
    \label{a-b=}
    \ee
Then substituting (\ref{n1=}) in (\ref{u-v}) gives
    \be
    \ft_1=\tan(\alpha+i\beta)=
    \frac{\tan\alpha+i\tanh\beta}{
    1-i\tan\alpha\tanh\beta}.
    \label{t1=2}
    \ee

Next, we use (\ref{n1=}), (\ref{t1=2}), and various trigonometric and hyperbolic identities to obtain the following explicit form of (\ref{eq3z}).
    \be
    \eta_1\sinh(2\beta)-\kappa_1\sin(2\alpha)+
    \left(\frac{\eta_1^2+\kappa_1^2+n_2^2}{n_2^2+1}\right)\cosh(2\beta)+
    \left(\frac{\eta_1^2+\kappa_1^2-n_2^2}{n_2^2+1}\right)\cos(2\alpha)\approx 0.
    \label{pert1}
    \ee
A similar analysis reveals the fact that the following quantity is an integer.
    \be
    \tilde m:=\gamma-\pi^{-1}{\rm arg}\left(\frac{1+i\cF}{1-i\cF}\right),
    \label{m-tilde=}
    \ee
where
    \bea
   \gamma&:=&\frac{4a_1n_2}{\lambda}+\xi-\nu,
   \label{gamma=}\\
   \xi&:=&\pi^{-1}\arctan\left\{
    \frac{2n_2[\eta_1\sin(2\alpha)+\kappa_1\sinh(2\beta)]}{
    [n_2^2-(\eta_1^2+\kappa_1^2)]\cosh(2\beta)-
    [n_2^2+\eta_1^2+\kappa_1^2]\cos(2\alpha)}\right\},
    \label{xi=}
   \eea
and ``arg$(z)$'' denotes the principal argument of $z$ that takes values in $(-\pi,\pi]$. The latter implies that $\tilde m$ is one of the two integers satisfying the condition:
    \be
    \gamma-1 \leq\tilde m<\gamma+1.
    \label{integer-part}
    \ee
A more important implication of (\ref{m-tilde=}) is that it leads to the following explicit form of (\ref{eq1z}).
    \be
    a_2-a_1\approx a_0+\frac{\lambda(2m-\tilde m)}{4n_2},
    \label{pert2}
    \ee
where
    \be
    a_0:=\frac{\lambda \xi}{4 n_2}.
    \label{a-zero=}
    \ee
Note that the ``$\arctan$'' appearing in (\ref{xi=}) stands for the principal value of ``$\tan^{-1}$'' that takes values in $[-\frac{\pi}{2},\frac{\pi}{2}]$. This in particular implies that $|\xi|\leq 1/2$. Therefore,
    \be
    |a_0|\leq\frac{\lambda}{8 n_2}.
    \label{bound-a}
    \ee
Another useful relation that follows from (\ref{gamma=}), (\ref{xi=}), and (\ref{a-zero=}) is
    \be
    \gamma=\frac{4n_2(a_1+a_0)}{\lambda}-\nu.
    \label{gamma=2}
    \ee

Eq.~(\ref{pert2}) is quite remarkable, for it indicates that if we choose $\eta_1,\kappa_1,n_2$ and $x_1=2\pi a_1/\lambda$ so that $x_1\gg 1$ and (\ref{pert1}) holds, then an OSS arises for a discrete set of values of the thickness of the coating. Because $2m-\tilde m$ is an integer, according to (\ref{pert2}) and (\ref{bound-a}),
    \be
    2m\gtrapprox \tilde m \geq\gamma-1,~~~~~~~~~~~
    a_2-a_1 \gtrapprox \frac{\lambda}{8 n_2}.
    \label{bound-2a}
    \ee
This means that the mode number $m$ and the thickness $a_2-a_1$ are bounded from below by $(\gamma-1)/2$ and $\lambda/(8n_2)$, respectively. Notice that these bounds only depend on $a_1$, $\lambda$ and $n_2$.

Next, we examine the physical implications of (\ref{pert1}) for a typical optically active material that satisfies
    \be
    |\kappa_1|\ll |\beta| \ll \eta_1\ll \alpha.
    \label{condi}
    \ee
In this case, we can ignore the terms of order two and higher in $\kappa_1$ and $\beta$ in our calculations. Implementing this approximation in (\ref{pert1}), using (\ref{a-b=}), and recalling that the imaginary part of $\fn_1$, i.e., $\kappa_1$, is related to the gain coefficient $g$ via $\kappa_1=-g\lambda/(4\pi)$, we find
    \be
    a_1g\approx \frac{\eta_1^2+n_2^2+(\eta_1^2-n_2^2)\cos(2\alpha)}{\eta_1(n_2^2+1)}.
    \label{a1=appr}
    \ee
This relation shows that the radius $a_1$ of the inner core is inversely proportional to the gain coefficient. Furthermore, it puts curious upper and lower bounds on the possible values of $a_1$ that are only sensitive to the gain coefficient and the real part of the refractive indices of the inner core and the outer shell:
    \be
    \frac{2 n_{\rm min}^2}{\eta_1g(n_2^2+1)}\lessapprox a_1
    \lessapprox \frac{2 n_{\rm max}^2}{\eta_1g(n_2^2+1)},
    \label{bound1}
    \ee
where  $n_{\rm max}$ and $n_{\rm min}$ are respectively the largest and smallest of $\eta_1$ and $n_2$.

In view of (\ref{bound1}), the larger $n_2$ is, the smaller the lower bound of $a_1$ gets. This confirms our expectation that coating a spherical gain medium by a material with higher refractive index reduces the lower bound on the radius of the gain medium. For example, if we take $n_2= 2.5$ and choose the inner core to be made of a dye gain material with
$\eta_1\approx 1.48$ and $g\approx 5~{\rm cm}^{-1}$, we find
    \be
    0.816~{\rm mm}\lessapprox a_1 \lessapprox 2.330~{\rm mm}.
    \label{a1-bound}
    \ee

Finally, we explore the consequences of (\ref{condi}). Implementing the above approximation scheme of neglecting second and higher order terms in $\beta$ and $\kappa_1$ in (\ref{pert2}) yields
    \be
    a_0\approx \frac{\lambda}{4\pi n_2}
    \arctan\left\{\frac{2\eta_1n_2\sin(2\alpha)}{n_2^2-\eta_1^2-(\eta_1^2+n_2^2)
    \cos(2\alpha)}\right\}.
    \label{a2-a1=appr}
    \ee
Taking $a_1=1~{\rm mm}$ to comply with (\ref{a1-bound}), choosing $\lambda=549~{\rm nm}$, $\eta_1\approx 1.48$, and $n_2=2.5$ as above, and using (\ref{pert2}), (\ref{gamma=}), (\ref{integer-part}),
(\ref{bound-2a}) and (\ref{a2-a1=appr}) we find
    \be
    a_2-a_1\approx [55~(2m-\tilde m)+27]~{\rm nm},~~~~~~~~
    \tilde m\approx \gamma =18122.
    \label{a2-a1-bound=2}
    \ee
The smallest allowed value of the thickness is therefore $a_2-a_1\approx a_0=27~{\rm nm}$.

In practice the thickness of the coating has a fixed value, and (\ref{a2-a1=appr}) and (\ref{a2-a1-bound=2}) determine an approximate value of the mode number $m$:
    \be
    m\approx\frac{2n_2(a_2-a_1-a_0)}{\lambda}+\frac{\tilde m}{2}.
    \label{m-approx}
    \ee
If we further approximate $\tilde m$ by $\gamma$ and use the expression (\ref{gamma=2}) for the latter, then (\ref{m-approx}) gives
    \be
    m\approx\frac{2n_2a_2}{\lambda}-\frac{\nu}{2}.
    \label{m-approx2}
    \ee
For example, for $a_2-a_1=5~\mu{\rm m}$, this relation gives $m\approx 9152$, which coincides with the numerical result obtained directly from (\ref{eq1z}).

We can also use (\ref{m-approx2}) to express the wavelength of the OSS in terms of the mode number. This gives
    \be
    \lambda\approx \frac{4n_2 a_2}{2m+\nu}.
    \label{lambda-approx}
    \ee
The analogous expression for the case that we remove the coating is given by Eq.~(33) of Ref.~\cite{pla-2011} and reads
     \be
    \lambda\approx \frac{4\eta_1 a_1}{2m+\nu+1}\approx \frac{4\eta_1 a_1}{2m+\nu}.
    \label{lambda-approx-z}
    \ee
Comparing (\ref{lambda-approx}) with (\ref{lambda-approx-z}) and noting that $n_2a_2>\eta_1a_1$, we can see that the presence of the coating increases the wavelength of the OSS associated for each mode number. Alternatively, it increases the value of the mode number for an OSS of given wavelength.

We conclude this section by pointing out that we can perform a perturbative calculation of OSS in a way that avoids the explicit appearance of the mode number $m$. According to (\ref{pert2}) and (\ref{a-zero=}), $\tan[4\pi n_2(a_2-a_1)/\lambda]=\tan(\pi\xi)$. In light of (\ref{xi=}) we can easily compute $\tan(\pi\xi)$ and express the latter equation in the form
    \bea
    &&\big\{(n_2^2+\eta_1+\kappa_1)
    \cos(2\alpha)-(n_2^2-\eta_1-\kappa_1)\cosh(2\beta)\big\}\sin[4\pi n_2\lambda^{-1}(a_2-a_1)]\nn\\
    &&\hspace{4cm}+2n_2\big\{\eta_1\sin(2\alpha)+\kappa_1\sinh(2\beta)\big\}
    \cos[4\pi n_2\lambda^{-1}(a_2-a_1)]=0.~~~~~
    \label{pert2-n}
    \eea
In summary, we have obtained the real equations (\ref{pert1})  and (\ref{pert2-n}) by imposing the complex equation (\ref{equation17}). We derived the latter  by performing first-order perturbation theory on (\ref{exactspect}) that determined OSS. It is important to note that every solution of (\ref{equation17}) is a solution of (\ref{pert1})  and (\ref{pert2-n}), but the converse may not be true. We have checked for some concrete examples and found that indeed this is the case. Therefore, we solve (\ref{pert1})  and (\ref{pert2-n}) by fixing all but two of the real parameters entering in these equations and then eliminate the solutions that violate (\ref{equation17}).

 \section{OSS of a Concrete Two-Layer Spherical Medium}

In general the refractive index $\fn$ of an optically active medium depends on the properties of the medium and the wavelength of the propagating EM wave. For example, for a gain medium that is obtained by doping a host medium of refraction index $n_0$ and modeled by a two-level atomic system with lower and upper level population densities $N_l$ and $N_u$, resonance frequency
$\omega_0$, and damping coefficient $\gamma$, it satisfies the dispersion relation  \cite{pra-2011a}:
    \be
    \fn^2= n_0^2-
    \frac{\hat\omega_p^2}{\hat\omega^2-1+i\hat\gamma\,\hat\omega},
    \label{epsilon}
    \ee
where $\hat\omega:=\omega/\omega_0$, $\hat\gamma:=\gamma/\omega_0$,
$\hat\omega_p:=(N_l-N_u)e^2/(m_e\varepsilon_0\omega_0^2)$, $e$ is
electron's charge, and $m_e$ is its mass. We can express $\hat\omega_p^2$ in terms of the imaginary part $\kappa_0$ of $\fn$ at the resonance
wavelength $\lambda_0:=2\pi c/\omega_0$ according to \cite{pra-2011a}
    \be
    \hat\omega_p^2\approx2n_0\hat\gamma\kappa_0,
    \label{plasma}
    \ee
where the approximation symbol means that we neglect quadratic and higher order terms in $\kappa_0$.

Inserting~(\ref{plasma}) in (\ref{epsilon}) and using
$\fn=\eta+i\kappa$, we obtain
    \be
    \eta\approx n_0+\kappa_0f_1(\hat\omega) ,~~~~
    \kappa\approx\kappa_0f_2(\hat\omega) ,
    \label{eqz301}
    \ee
where
    \be
    f_1(\hat\omega):=\frac{\hat\gamma(1-\hat\omega^2)}{(1-\hat\omega^2)^2+
    \hat\gamma^2\hat\omega^2},~~
    f_2(\hat\omega):=\frac{\hat\gamma^2\hat\omega}{(1-\hat\omega^2)^2+
    \hat\gamma^2\hat\omega^2}.
    \label{fs}
    \ee
We also note that the gain coefficient of such a medium is given by $g = -4\pi \kappa / \lambda$. In particular, we can use this relation to express $\kappa_0$ in terms of the gain coefficient $g_0$ at the resonance wavelength $\lambda_0$ according to
    \be
    \kappa_0=-\frac{ \lambda_{0}g_0}{4\pi}.
    \label{k-zero=z}
    \ee
In the following, we employ  (\ref{eqz301}) -- (\ref{k-zero=z}) to parameterize the refractive indices $\fn_{1}$ and $\fn_{2}$ that enter in the description of our two-layer spherical model.

In order to explore the effect of the outer shell on the behavior of spectral singularities, we consider a spherical dye laser medium confined in a thin spherical shell of higher-refractive index glass. We suppose that the refractive index of the glass takes a constant real value and parameterize the location of the spectral singularities using the resonance gain coefficient, $g_{0}$,  of the dye and the wavelength $\lambda$.

\begin{figure}[h]
    \begin{center}
    \epsfig{file=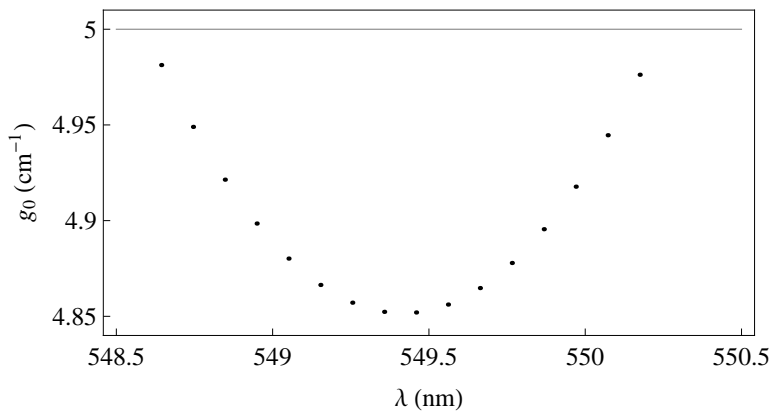,scale=0.80}\vspace{-.4cm}
    \caption{Spectral singularities of the Rose Bengal-DMSO dye gain  medium (\ref{specific}) confined in a spherical glass shell of outer radius 1 mm, thickness 5 $\mu{\rm m}$, and refractive index 2.5. The minimum corresponds to $\lambda=\lambda^{(1)}\approx 549.459~{\rm nm}$ that is larger than the resonance wavelength $\lambda_0=549~{\rm nm}$. The grey horizontal line represents the experimental upper bound on $g_0$.} \label{figure2}
    \end{center}
    \end{figure}

Consider confining a Rose Bengal-DMSO (Dimethyl sulfoxide) solution with characteristics \cite{silfvast,Rose}
    \be
    n_0=1.479,~~~\lambda_0=549\,{\rm nm},~~~
    \hat\gamma=0.062,~~~g_0\leq 5\,{\rm cm}^{-1},
    \label{specific}
    \ee
in a spherical glass shell of outer radius $a_2=1~{\rm mm}$, thickness $a_2-a_1=5~\mu{\rm m}$, and refractive index $n_2 = 2.5$.\footnote{For the details of developing such a high-refractive index glass, see
\cite{ref25,highindex}.} Figure~\ref{figure2} and Table~\ref{table-1} show the results of our numerical calculation of the location of spectral singularities that use (\ref{exactspect}).
Table~\ref{table-1} also gives the results of our perturbative calculations and demonstrates their good agreement with the numerical results.
    \begin{table}
    \begin{center}
    \begin{tabular}{|c|c|c|c|c|}
      \hline
      $\ell$ & $m$ & $g_{0}^{(\ell)} ({\rm cm}^{-1})$  & $\lambda_{\rm pert.}^{(\ell)} (nm)$ & $\lambda_{\rm exact}^{(\ell)} ({\rm nm})$
      \\ \hline
      1 & 9099 & 4.8520 & 549.458833 & 549.458836 \\
      2 & 9101 & 4.8523 & 549.356779 & 549.356781 \\
      3 & 9098 & 4.8561 & 549.560925 & 549.560927 \\
      4 & 9103 & 4.8571 & 549.254762 & 549.254765 \\
      5 & 9096 & 4.8647 & 549.663054 & 549.663056 \\
      6 & 9104 & 4.8664 & 549.152785 & 549.152787 \\
      7 & 9094 & 4.8779 & 549.765220 & 549.765222 \\
      \hline
     \end{tabular}
     \vspace{-.1cm}
     \caption{The mode number $m$, gain coefficient $g_0$, and wavelength $\lambda^{(\ell)}$ for the spectral singularities of the Rose Bengal-DMSO dye gain  medium (\ref{specific}) confined in a spherical glass shell of outer radius 1 mm, thickness 5 $\mu{\rm m}$, and refractive index 2.5. The subscripts  ``exact'' and  ``pert.'' refer to the results of numerical and perturbative calculations, respectively. These calculations give the same values for $g_0$ up to six significant figures.} \label{table-1}
     \end{center}
     \end{table}

Let $g_0^{(1)}$ denote the smallest value of the gain coefficient $g_0$ that is capable of producing an OSS, and $\lambda^{(1)}$ be the wavelength of this OSS.  We recall from Ref.~\cite{pla-2011} that in the absence of coating $\lambda^{(1)}$ is essentially identical with the resonance wavelength $\lambda_0$. According to Figure~\ref{figure2} and Table~\ref{table-1},  the presence of the glass coating causes $\lambda^{(1)}$ to be slightly red-shifted.

\begin{figure}[h!]
    \begin{center}
    \epsfig{file=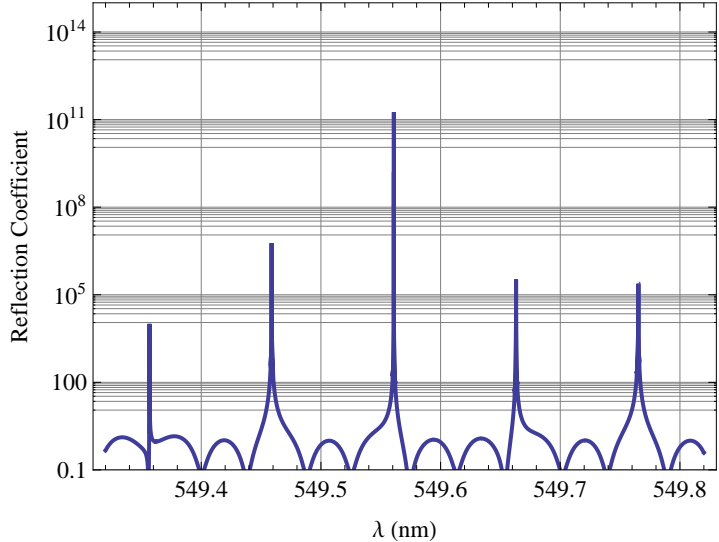,scale=0.75}\vspace{-.4cm}
    \caption{(Color online) Graph of reflection coefficient for the outer radius of
    1~$mm$ and thickness of 5 $\mu m$ when the gain coefficient is
    $g_{0}^{(3)} = $ 4.85614735 $cm^{-1}$. The central peak represents an OSS.}\label{figure10}
    \end{center}
    \end{figure}

Figure~\ref{figure10} shows a logarithmic plot of the reflection coefficient as a function of the wavelength for the sample considered in Figure~\ref{figure2} and Table~\ref{table-1} with $g_0=g_{0}^{(3)}=4.85614735~{\rm cm}^{-1}$. In this case, there is an OSS at $\lambda=\lambda^{(3)}=549.56092702~{\rm nm}$ that corresponds to the central peak in Figure~\ref{figure10}. Unlike the other peaks shown in this figure, the hight of the central peak increases indefinitely as we use more and more accurate numerical values for the parameters of the system. This is a clear indication that it corresponds to a spectral singularity.

\begin{figure}[h]
    \begin{center}
    \epsfig{file=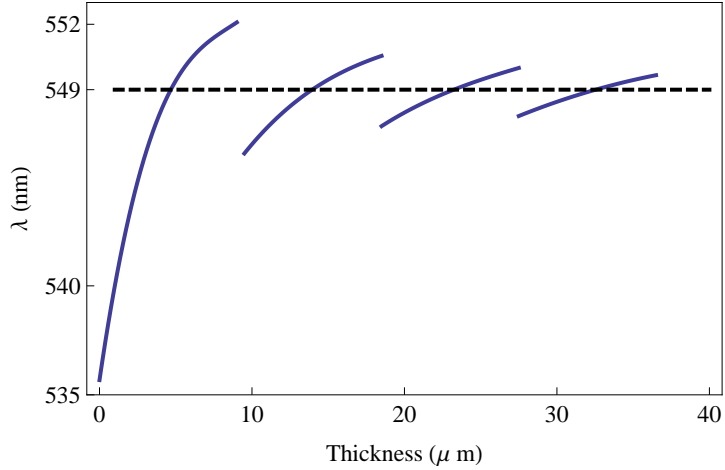,scale=0.75}\vspace{-.4cm}
    \caption{(Color online) Graph of the wavelength $\lambda^{(1)}$ as a function of the thickness of coating. The inner radius is held at $a_1=1.6~{\rm
    mm}$. As we change the thickness, $\lambda^{(1)}$ undergoes an
    infinite set of jumps that oscillate about $\lambda_0$ with a
    decreasing amplitude.}\label{fig3}
    \end{center}
    \end{figure}

Next, we explore the effect of changing the thickness of coating $a_2-a_1$ on the $\lambda^{(1)}$ while keeping the inner radius fixed. Figure~\ref{fig3} shows the graph of $\lambda^{(1)}$ as a function of thickness. As seen from this figure, as we increase the thickness, $\lambda^{(1)}$ undergos an infinite set of jumps that oscillate about the resonance wavelength $\lambda_0$ with a decreasing amplitude. Our numerical results show that unlike the wavelength $\lambda^{(1)}$, the gain coefficient $g_0^{(1)}$ does not experience a noticeable change due to an increase in the thickness.

\begin{figure}[h]
    \begin{center}
    \epsfig{file=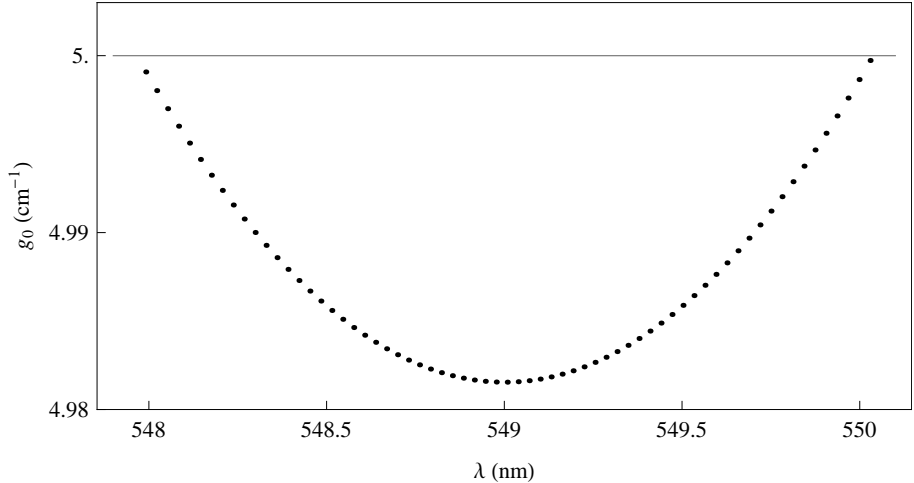,scale=0.95}\vspace{-.4cm}
    \caption{OSS of an uncoated spherical Rose Bengal-DMSO dye gain medium with specifications (\ref{specific}) and radius 3.3~mm. The minimum gain coefficient necessary for generating an OSS and the corresponding wavelength are respectively 4.9815~${\rm cm}^{-1}$ and 549.008~nm. The horizontal grey line represents the experimental upper bound on $g_0$. There are 66 OSS complying with this bound.}
    \label{figure4}
    \end{center}
    \end{figure}

\begin{figure}[h]
    \begin{center}
    \epsfig{file=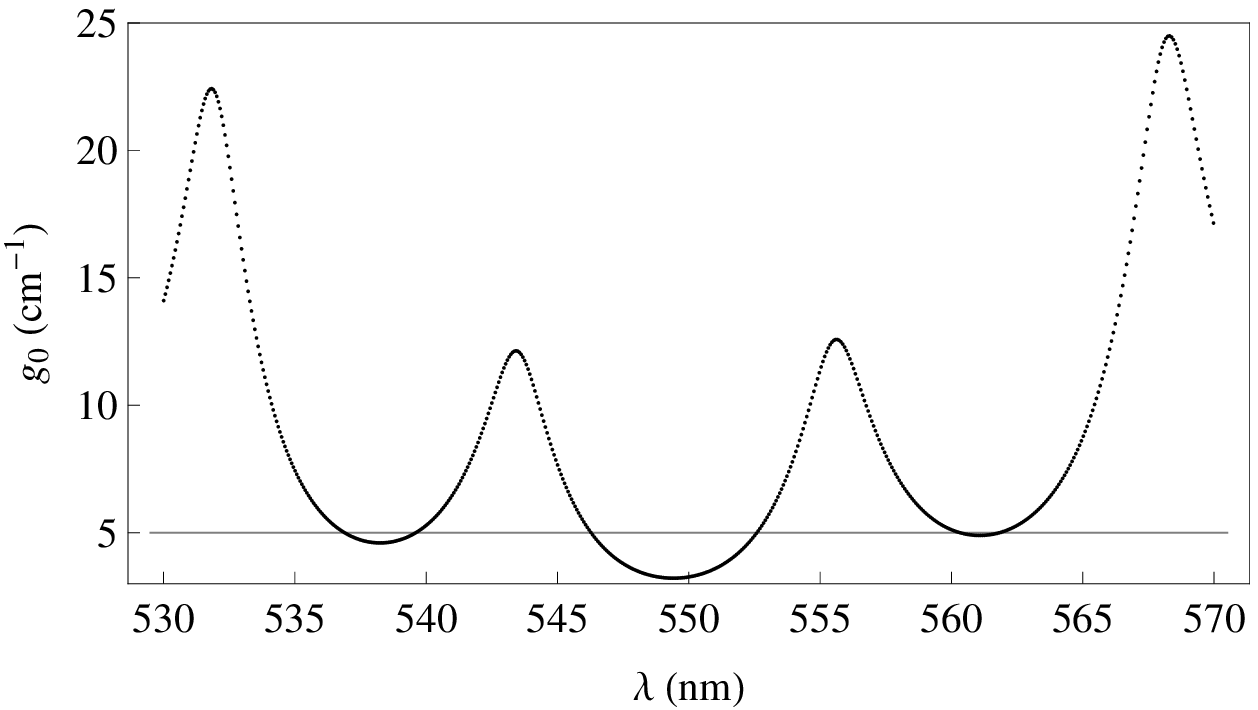,scale=0.75}\vspace{-.4cm}
    \caption{OSS of a coated spherical Rose Bengal-DMSO dye gain medium with specifications (\ref{specific}), inner radius 1.5~mm, and coating thickness 5~$\mu$m. The horizontal grey line represents the experimental upper bound on $g_0$. There are three groups of OSS below this bound. They appear in the spectral ranges 536.941436-539.544783~nm, 546.335226-552.571190~nm, and 560.400334-561.815048~nm and contain 41, 93, and 21 members, respectively. }\label{figure5}
    \end{center}
    \end{figure}

\begin{figure}[h!]
    \begin{center}
    \epsfig{file=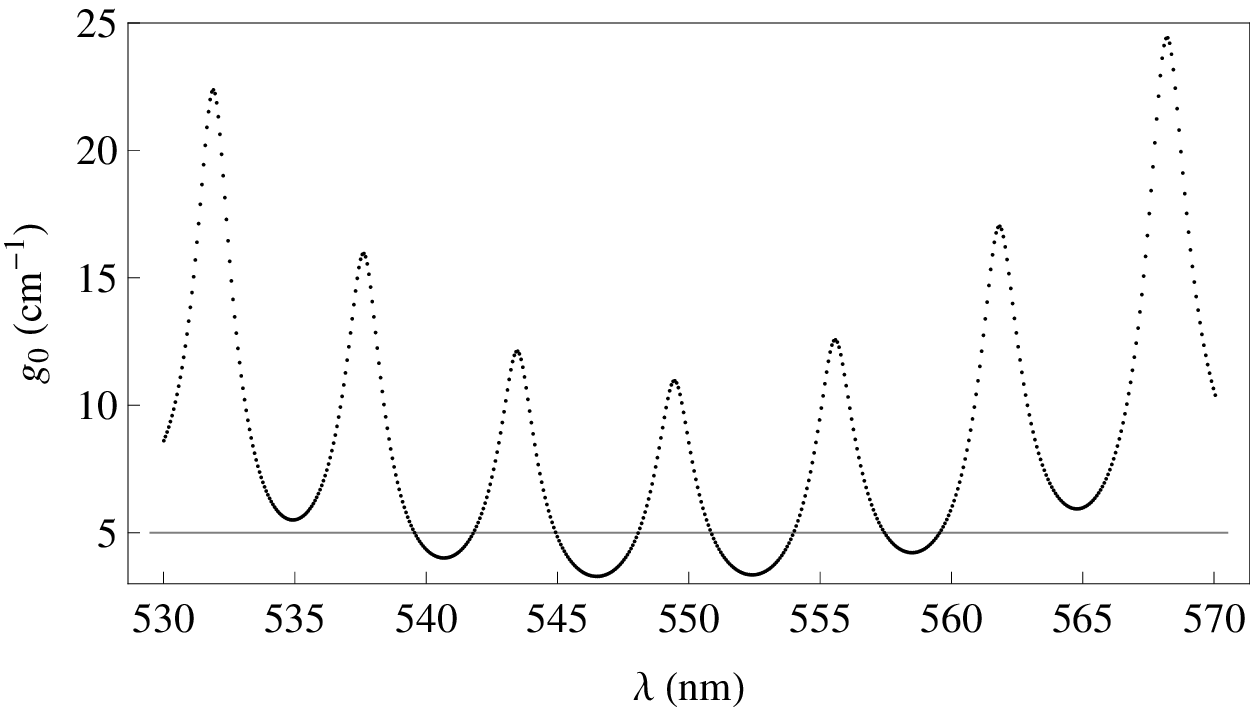,scale=0.75}\vspace{-.4cm}
    \caption{OSS of a coated spherical Rose Bengal-DMSO dye gain medium with specifications (\ref{specific}), inner radius 1.5~mm, and coating thickness 10~$\mu$m. There are four groups of OSS fulfilling the experimental upper bound on $g_0$ (the horizontal grey line.) They appear in the spectral ranges 539.621948-541.782279~nm, 544.992898-548.067753~nm, 550.878022-553.951299~nm, and 557.516880-559.542755~nm and contain 34, 47, 46, and 30 members, respectively.}\label{figure6}
    \end{center}
    \end{figure}

Figure~\ref{figure4} shows the location of OSS for a spherical dye gain medium without a coating. Comparing this figure with Figure~\ref{figure2}, we see that except for the shift in the value of $\lambda^{(1)}$, the distribution of the OSS in the $g_0$-$\lambda$ plane does not seem to get affected by the presence of the coating glass.
Exploring a wider range of values of $\lambda$ and $g_0$ reveals a different picture. Figures~\ref{figure5} and \ref{figure6} show the location of  OSS in a wider spectral range for a coated spherical Rose Bengal-DMSO sample with specifics  (\ref{specific}),  inner radius $1.5~{\rm mm}$, and two different values of the coating thickness, namely $5~\mu{\rm m}$ and $10~\mu{\rm m}$. As these figures show, OSS are located on curves with multiple local minima. The number of these minima that fulfill the experimental upper bound of $g_0\leq 5~{\rm cm}^{-1}$ is an increasing function of the thickness of the coating. For $a_2-a_1=5~\mu{\rm m}$ (Figure~\ref{figure5}), this number is three, i.e., there are three distinct groups of OSS satisfying the bound: $g_0\leq 5~{\rm cm}^{-1}$. These appear in the spectral ranges 536.941436-539.544783~nm, 546.335226-552.571190~nm, and 560.400334-561.815048~nm and respectively contain 41, 93, and 21 members. Their central member that corresponds to the smallest gain coefficient appear at $(\lambda,g_0)=(538.239847,4.601744), (549.435560,3.218290)$, and $(561.106835,4.895167)$ in $({\rm nm},{\rm cm}^{-1})$ units. For $a_2-a_1=10~\mu{\rm m}$ (Figure~\ref{figure6}), there are four groups of OSS satisfying the bound on $g_0$. They include 34, 47, 46, and 30 members with wavelengths ranging over 539.621948-541.782279~nm, 544.992898-548.067753~nm, 550.878022-553.951299~nm, and 557.516880-559.542755~nm, respectively. The central members of these four groups that correspond to the local minima of $g_0$ have $(\lambda,g_0)$ values: $(540.666848, 4.010267)$, $(546.525852,3.287748)$, $(552.444850,3.348007)$ and $(558.493430,4.217539)$ in $({\rm nm},{\rm cm}^{-1})$ units.
In particular, the presence of the glass coating not only allows for generating OSS in a smaller gain medium, but it produces spectral gaps in the spectral range within which these OSS are located. These remarkable observations should in principle be verifiable experimentally.

\begin{figure}[h!]
    \begin{center}
    \epsfig{file=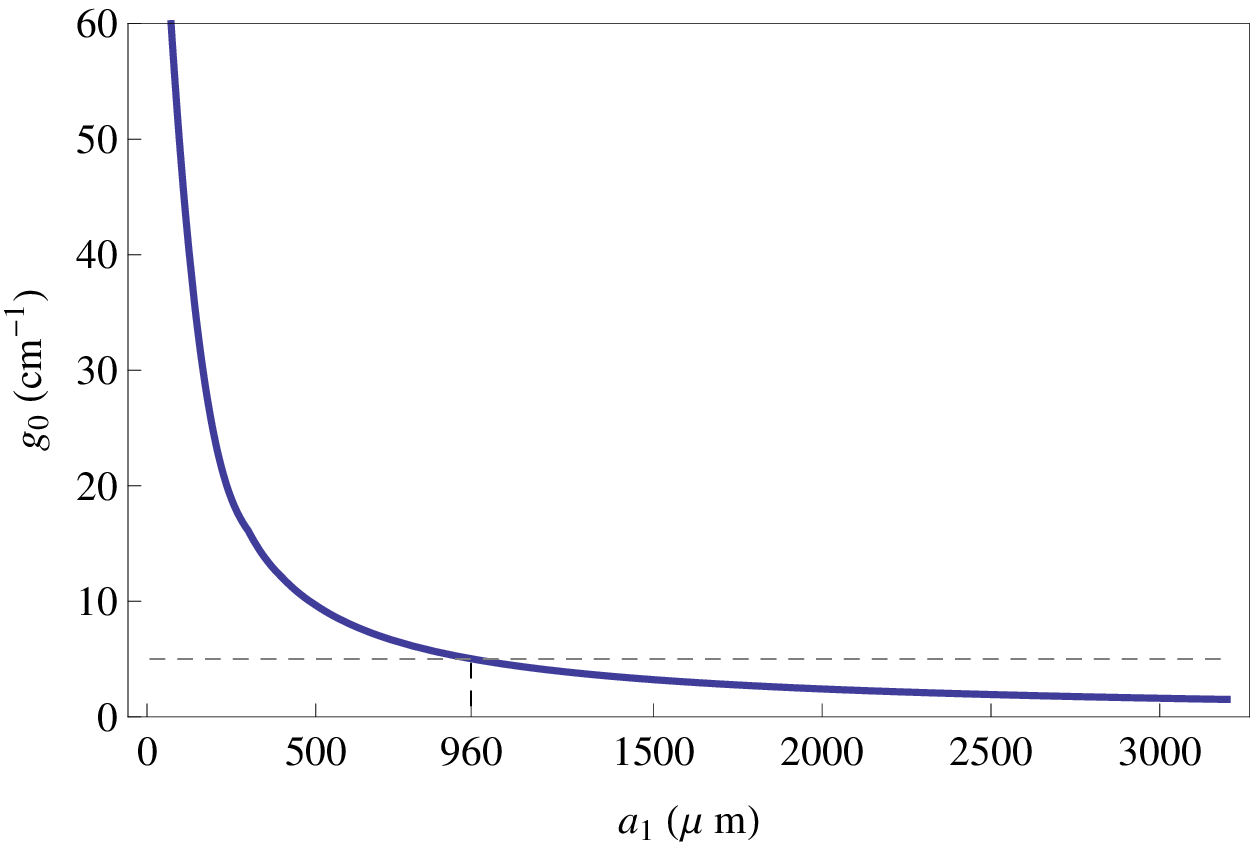,scale=0.75}\vspace{-.4cm}
    \caption{(Color online) Graph of $g_0$ as a function of the inner radius $a_1$. The coating thickness is kept fixed at 5~$\mu m$. Dashed line corresponds to the experimental bound $g_0\leq 5 {\rm cm}^{-1}$. For $g_0=5~{\rm cm}^{-1}$ the smallest value of the inner radius that supports an OSS is about 960~$\mu m$. For larger values of $a_1$, there are OSS with $g_0<5~{\rm cm}^{-1}$.
    }\label{figure7}
    \end{center}
    \end{figure}

\begin{figure}[h!]
    \begin{center}
    \epsfig{file=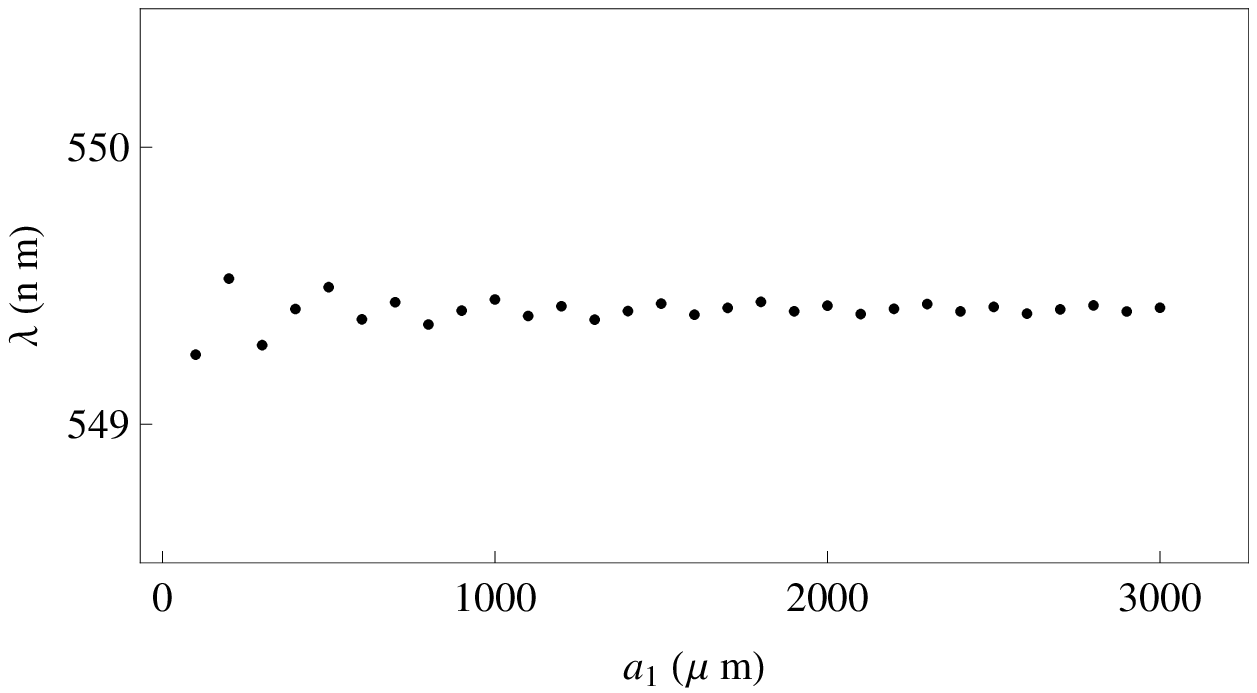,scale=0.75}\vspace{-.4cm}
    \caption{Effect of changing the inner radius on the wavelength $\lambda^{(1)}$ for the coating thickness of 5~$\mu$m. The displayed dots correspond to increments of the radius $a_1$ of 100~$\mu$m. For continuous variations of the radius these dots fluctuate around the line $\lambda^{(1)}=549.4~{\rm nm}$ with a decreasing amplitude as the gain coefficient increases.}\label{figure8}
    \end{center}
    \end{figure}

Next, we study the effect of changing the radius of the inner core on the location of OSS for a fixed value of the coating  thickness. Figure~\ref{figure7} shows the graph of the minimum gain coefficient $g_0$ necessary for generating an OSS as a function of $a_1$ for the coated spherical Rose Bengal-DMSO dye gain medium with specifications (\ref{specific}) and thickness $5~\mu{\rm m}$. For $a_1\leq a_1^{(\rm min)}\approx 0.96~{\rm mm}$ no OSS can be created. Recalling that for an uncoated sample $a_1^{(\rm min)}\approx 3.3~{\rm mm}$, this corresponds to a three-fold decrease in the size of the gain medium. As we increase $a_1$ starting from the critical value $0.96~{\rm mm}$, the minimum gain coefficient necessary for generating an OSS, namely $g_0^{(1)}$ decreases. Both of these observations are in agreement with our perturbative results. Moreover, it turns out that for fixed values of the coating thickness the wavelength $\lambda^{(1)}$ is not sensitive to the variations of the inner radius.  Figure~\ref{figure8} demonstrates this behavior.
As our perturbative treatment shows the larger the refractive index of coating is the smaller the minimum radius $a_1^{(\rm min)}$ gets. For example if we use a glass coating with the same thickness ($5~\mu{\rm m}$) but a slightly lower index of refraction, say $n_2=1.93$ as in \cite{ref25}, the value of $a_1^{(\rm min)}$ increases by a factor of 2  (to about $1.8~{\rm mm}$.)

\begin{figure}[h!]
    \begin{center}
    \epsfig{file=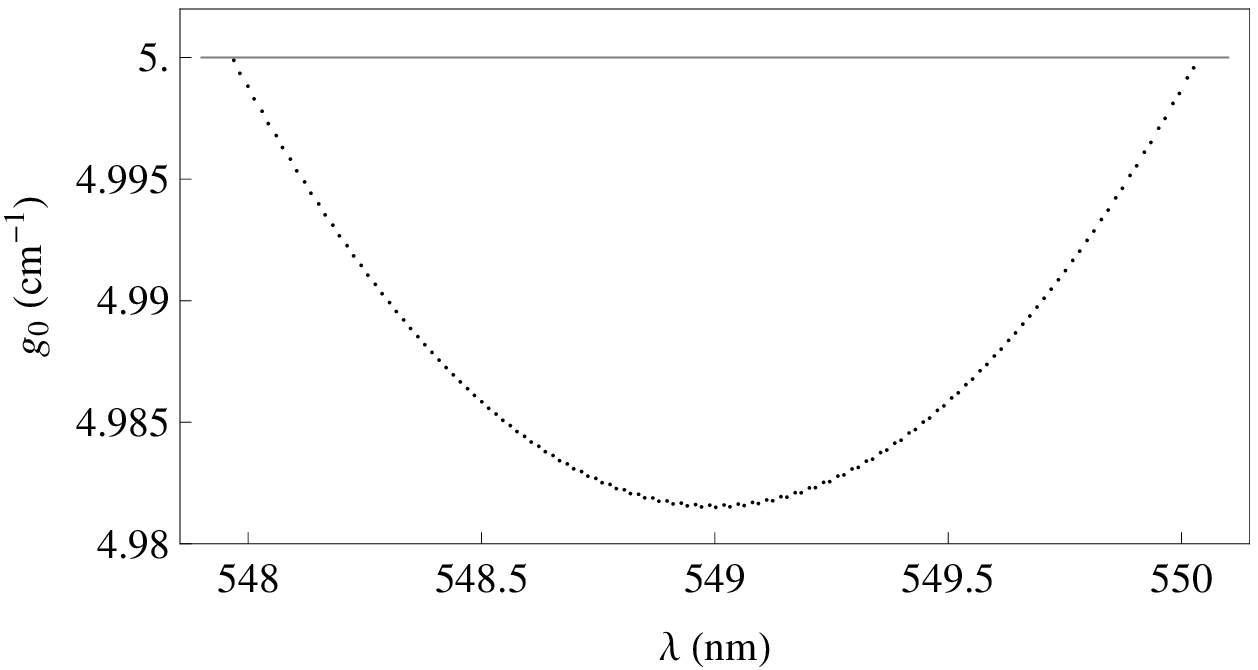,scale=0.85}\vspace{-.4cm}
    \caption{OSS for a double-layer sphere consisting of a glass core with
    refractive index $\fn_{1}=2.5$ and an outer shell filled with a Rose Bengal-DMSO dye solution with specifics~(\ref{specific}). The inner and outer radii are taken as $a_1=2~{\rm mm}$ and $a_2=5.3~{\rm mm}$. The grey line represents the experimental upper bound on $g_0$. There are 136 OSS fulfilling this bound. They are located in the spectral range: 547.969294-550.025813~nm. The OSS with minimum gain has coordinates $\lambda^{(1)} = 549.001882~{\rm nm}$ and $g_{0}^{(1)} = 4.981501~{\rm cm}^{-1}$.}\label{figure11}
    \end{center}
    \end{figure}

We have also investigated the formation of OSS in a two-layer
spherical model in which the inner core is made of a
higher-refractive index glass and the outer shell is filled with a
dye gain material. Using the same glass and gain material and taking
the core radius to be 2~mm, we found that the minimum shell
thickness required for producing an OSS was about 3.28~mm. This
coincides with the minimum radius supporting an OSS for a sphere
filled with the same gain material. Figure~\ref{figure11} shows the
location of OSS for the thickness of 3.3~mm as given by
(\ref{exactspect}). The main difference between this figure and
Figure~\ref{figure4} is that the number of OSS has doubled. To our
knowledge this is the only effect of the presence of the glass core.
We also find that using a glass core with a lower index of
refraction does not have a sizable effect on the location and number
of OSS.

\section{Concluding Remarks}

A gain medium can emit electromagnetic waves provided that we adjust
its parameters so that an optical spectral singularity is created.
The time-reversal of  this phenomenon corresponds to the (coherent
perfect) absorption of electromagnetic waves. This is the
theoretical basis of antilasing. The simplest example of a gain
medium that is localized in space and is capable of realizing an
optical spectral singularity is a spherical gain medium. It supports
spectral singularities in a radial transverse mode provided that its
radius exceeds a critical value $a^{(\rm min)}$.

For the typical dye laser material that we consider in
Ref.~\cite{pla-2011}, $a^{(\rm min)}\approx 3.28$~mm. In practice
maintaining a uniform gain coefficient within a spherical sample of
this size can be difficult. The main motivation for the current
study is the idea that coating a spherical gain medium by a
high-refractive index material can reduce $a^{(\rm min)}$. We have
shown by explicit perturbative and numerical calculations that this
is actually the case. In particular, coating a spherical active dye
medium with a glass of thickness 5~$\mu$m and index of refractive
$2.5$ reduces $a^{(\rm min)}$ by about a factor of 3. Furthermore,
we have found that the presence of coating leads to a much richer
structure as far as the location of spectral singularities are
concerned. In particular, the coating causes a small shift in the
wavelength of the spectral singularity that requires the least
amount of gain. More importantly, it produces a clustering of
spectral singularities into groups separated by sizable spectral
gaps. This behavior may find an application in producing tunable
lasers with rather wide spectral gaps. Another interesting
possibility is to generate laser pulses in different spectral ranges
by periodically altering the gain coefficients of the active core.
For example, as shown in Figures~\ref{figure5} and \ref{figure7},
increasing the gain coefficient so that it passes one or more of the
local minima of the OSS curves leads to lasing in two or more
spectral ranges that are separated by gaps of several nm in
width.\footnote{For an experimental realization of the OSSs we have
examined and their possible applications, one may try to use the SHG
of a diode pumped Nd-YAG (continuous wave) laser at 532 nm
wavelength in the plane-wave configuration. To achieve reasonably
uniform gain, one probably needs to use multiple pumping. The
details and a discussion of alternative pumping methods are beyond
the scope of the present paper and the expertise of the authors.}

If we use a glass spherical core and an active dye outer shell, the
thickness of the shell required to induce an optical spectral
singularity is about the $a^{(\rm min)}$ for an uncoated spherical
gain medium consisting of the same dye. The presence of the glass
core does not seem to have a significant effect except for the
doubling of the number of the spectral singularities. We plan to
explore the reasons for this phenomenon in a more general context.

The results we have reported apply for a TEM electromagnetic wave propagating in the radial direction. As we show in the appendix, we can extend our investigation to study optical spectral singularities for electromagnetic waves in the TE and TM modes. It turns out that the results are similar to those for the TEM mode. For example the size of the $a^{\rm min}$ remains essentially the same.

The emergence of an optical spectral singularity is a common feature of any lasing system. This includes typical coated or uncoated spherical lasers that can be realized using much smaller, micron size dye samples \cite{ref25,Vitaly}. The reason is that these lasers involve exciting whispering gallery modes \cite{Matsko}. We intend to conduct a through study of optical spectral singularities for whispering gallery modes.

\vspace{.5cm} \noindent {\em \textbf{Acknowledgments:}} We wish to express our gratitude to Ali Serpeng\"{u}zel for reading the draft of this article and making many useful comments and suggestions. This work has been
supported by  the Scientific and Technological Research Council of
Turkey (T\"UB\.{I}TAK) in the framework of the project no: 110T611
and the Turkish Academy of Sciences (T\"UBA).

\section*{Appendix}

In this appendix we extend our analysis of optical spectral
singularities to the general TE($\ell$, $m$)  \linebreak  and
TM($\ell$, $m$) modes of our two-layer spherical system.

We begin our treatment by writing  the solutions of the Maxwell's
equations in TE and TM modes of a spherical system as follows
\cite{jackson}.
    \begin{align}
        &\vec{E} = Z_{0} \sum_{\ell, m} \left[\frac{i}{k \fz(r)} a_{M} (\ell, m) \vec{\nabla} \times f_{\ell} (\tilde{k} r) \vec{X}_{\ell m} + a_{E} (\ell, m) g_{\ell} (\tilde{k} r)  \vec{X}_{\ell
        m}\right] \label{equation76}\\
        &\vec{H} = \sum_{\ell,m} \left[a_{M} (\ell, m) f_{\ell} (\tilde{k} r) \vec{X}_{\ell m} - \frac{i}{k} a_{E} (\ell, m) \vec{\nabla} \times g_{\ell} (\tilde{k} r)  \vec{X}_{\ell m}\right]
        \label{equation77}
    \end{align}
where $\ell$ and $m$ take integer values in ranges $[0, \infty)$ and
$[-\ell, \ell]$ respectively, $a_{E}$ and $a_{M}$ are the
coefficients of TE and TM modes respectively, $Z_{0} :=
\sqrt{\mu_{0}/\epsilon_{0}}$ is the impedance of the vacuum,
$\fz(r)$ is given in (\ref{e1}),  both $f_{\ell}(\tilde{k} r)$ and
$g_{\ell}(\tilde{k} r)$ have form
    \be
    f_{\ell} (\tilde{k} r), g_{\ell} (\tilde{k} r)=\left\{\begin{array}{ccc}
    A_1 j_\nu(k_1r) + B_1 n_\nu (k_1r) &{\rm for}& r<a_1,\\
    A_2 j_\nu(k_2r) + B_2 n_\nu (k_2r) &{\rm for}& a_1<r<a_2,\\
    A_3 h_\nu^{(1)} (kr) + B_3 h_\nu^{(2)}(kr)&{\rm for}&  r>a_2,\end{array}\right.
    \label{78}
    \ee
$\vec{X}_{\ell m} (\theta, \phi) := \frac{1}{\sqrt{\ell (\ell + 1)}} \,\vec{L} Y_{\ell , m}$, $\vec{L}:=-i\vec{r} \times \vec{\nabla}$,
$ Y_{\ell , m}$ are spherical harmonics:
     \be
      Y_{\ell , m} (\theta, \phi):= \sqrt{\frac{2\ell + 1}{4\pi}\frac{(\ell - m)!}{(\ell +      m)!}} P_{\ell}^{m} (\cos \theta) e^{i m \phi}, \notag
     \ee
and $P_{\ell}^{m}$ are the associated Legendre functions. Note that
$\vec{X}_{\ell m}$ fulfil the orthogonality relations: $\int \vec{X}_{\ell ' m'}^{\ast} . \vec{X}_{\ell m} d\Omega = \delta_{\ell \ell'} \delta_{m m'}$ and $\int \vec{X}_{\ell ' m'}^{\ast} . (\vec{r} \times \vec{X}_{\ell m}) d\Omega = 0$, where $d\Omega := \sin \theta d\theta d\phi$, and being proportion to the angular momentum operator, $\vec L$ satisfies $L^{2} Y_{\ell m} = \ell (\ell + 1) Y_{\ell m}$.

Expressing the electric and magnetic fields in terms of their
components in the spherical coordinates, we have $\vec{E}= \hat{r} E_{r} (r, \theta, \phi) + \hat{\theta} E_{\theta} (r, \theta, \phi) + \hat{\phi} E_{\phi} (r, \theta, \phi)$ and $\vec{H}= \hat{r}H_{r} (r, \theta, \phi) + \hat{\theta} H_{\theta} (r, \theta, \phi) + \hat{\phi} H_{\phi} (r, \theta, \phi)$, where
    \begin{align}
    E_{r} &= Z_{0} \sum_{\ell, m}
    \frac{a_{E}(\ell, m) b(\ell, m)f_{\ell} (\tilde{k} r)}{kr
    \fz(r)}\left[\frac{\partial \Pi_{\ell, m} (\theta)}{\partial \theta} - m^{2} \frac{\Omega_{\ell,m}(\theta)}{\sin \theta}\right]  e^{i m \phi}\notag\\
    E_{\theta} &= - Z_{0} \sum_{\ell, m} b(\ell, m)
    \left[\frac{a_{E}(\ell, m)}{k \fz(r)} \frac{\partial f_{\ell} (\tilde{k} r)}{\partial r} \Pi_{\ell, m} (\theta) + m a_{M}(\ell, m) g_{\ell} (\tilde{k} r) \Omega_{\ell,m}(\theta)\right] e^{i m \phi}\notag\\
    E_{\phi} &= - i Z_{0} \sum_{\ell, m} b(\ell, m)
    \left[\frac{m a_{E}(\ell, m)}{k \fz(r)} \frac{\partial f_{\ell} (\tilde{k} r)}{\partial r} \Omega_{\ell,m}(\theta) + a_{M}(\ell, m) g_{\ell} (\tilde{k} r) \Pi_{\ell, m} (\theta)\right] e^{i m \phi} \notag\\
    H_{r} &= - \sum_{\ell, m} \frac{a_{M}(\ell, m) b(\ell, m)g_{\ell} (\tilde{kr})}{kr} \left[\frac{\partial \Pi_{\ell, m} (\theta)}{\partial \theta} - m^{2} \frac{\Omega_{\ell, m}(\theta)}{\sin \theta}\right] e^{i m \phi} \notag\\
    H_{\theta} &= \sum_{\ell, m} b(\ell, m)  \left[- m a_{E}(\ell, m) f_{\ell} (\tilde{k} r)\Omega_{\ell, m}(\theta) + \frac{a_{M}(\ell, m)}{k} \frac{\partial g_{\ell} (\tilde{k} r)}{\partial r} \Pi_{\ell, m} (\theta)\right]e^{i m \phi} \notag\\
    H_{\phi} &= i \sum_{\ell, m} b(\ell, m) \left[- a_{E}(\ell, m) f_{\ell} (\tilde{k} r)\Pi_{\ell, m} (\theta) + \frac{m a_{M}(\ell, m)}{k} \frac{\partial g_{\ell} (\tilde{k} r)}{\partial r} \Omega_{\ell,
    m}(\theta)\right]e^{i m \phi}  \notag\\
    b(\ell, m)& := \frac{1}{\ell (\ell +1)}\sqrt{\frac{2\ell + 1}{4\pi}
     \frac{(\ell - m)!}{(\ell + m)!}},~~~
    \Pi_{\ell, m} (\theta) := \frac{\partial}{\partial \theta}
     P_{\ell}^{m} (\cos \theta),~~~\Omega_{\ell, m}(\theta) :=
    \frac{P_{\ell}^{m} (\cos \theta)}{\sin \theta}.\notag
    \end{align}
Imposing the physical matching conditions at $r = a_1$ and $r=a_2$,
we find that the tangential components of $\vec{E}$ and $\vec{H}$
are continuous along the boundaries, i.e., $E_{\theta}^{in} = E_{\theta}^{out}$, $E_{\phi}^{in} =E_{\phi}^{out}$, $H_{\theta}^{in} = H_{\theta}^{out}$, and $H_{\phi}^{in} = H_{\phi}^{out}$.

For TE modes, we can express these boundary conditions as
    \be
   \mathbf{K}_{11}^{E}\left[\begin{array}{c}
    a_{E1} \\
    0\end{array}\right]= \mathbf{K}_{12}^{E}\left[\begin{array}{c}
    a_{E2} \\
    b_{E2}\end{array}\right],~~~~~~
   \mathbf{K}_{22}^{E} \left[\begin{array}{c}
    a_{E2} \\
    b_{E2} \end{array}\right]=
    \mathbf{L}^{E} \left[\begin{array}{c}
    a_{E3} \\
    b_{E3} \end{array}\right],
    \label{e82}
    \ee
where for all $p,q=1,2$,
    \be
    \mathbf{K}_{p q}^{E}:= \left[
    \begin{array}{cc}
    j_{\ell} (k_{q}a_p) & n_{\ell} (k_{q}a_p) \\
    \tilde j_{\ell} (k_{q}a_p) & \tilde n_{\ell} (k_{q}a_p)
    \end{array}\right],~~~~~~
    \mathbf{L}^{E}:= \left[
    \begin{array}{cc}
    h_{\ell}^{(1)} (ka_2) & h_{\ell}^{(2)} (ka_2) \\
    \tilde h_{\ell}^{(1)} (ka_2) & \tilde h_{\ell}^{(2)} (ka_2)
    \end{array}\right].
    \label{KL83}
    \ee
Note that these are respectively  identical with (\ref{e7}) and
(\ref{KL}), i.e., $\mathbf{K}_{p q}^{E}=\mathbf{K}_{p q}$ and
$\mathbf{L}^{E}=\mathbf{L}$. Therefore, as far as the study of the spectral singularities are concerned, we obtain similar results except that $\nu =\sqrt{5}/2$ is now replaced by $\ell$. As we see from (\ref{e41}), this does not influence the parameters of the system. It only changes the mode numbers associated with spectral singularities.

Similarly, for TM modes, we find the following set of boundary
conditions.
    \be
   \mathbf{K}_{11}^{M}\left[\begin{array}{c}
    a_{M1} \\
    0\end{array}\right]= \mathbf{K}_{12}^{M}\left[\begin{array}{c}
    a_{M2} \\
    b_{M2}\end{array}\right],~~~~~~
   \mathbf{K}_{22}^{M} \left[\begin{array}{c}
    a_{M2} \\
    b_{M2} \end{array}\right]=
    \mathbf{L}\left[\begin{array}{c}
    a_{M3} \\
    b_{M3} \end{array}\right],
    \ee
where, for all $p, q=1,2$,
    \be
    \mathbf{K}_{p q}^{M}:= \left[
    \begin{array}{cc}
    j_{\ell} (k_{q}a_p) & n_{\ell} (k_{q}a_p) \\
    \displaystyle\frac{\tilde j_{\ell} (k_{q}a_p)}{\fn_{q}^{2}} &
     \displaystyle\frac{\tilde n_{\ell}
    (k_{q}a_p)}{\fn_{q}^{2}}
    \end{array}\right].
    \ee
As we see the only difference between $\mathbf{K}_{p q}^{M}$ and
$\mathbf{K}_{p q}$ is the appearance of the factor $1/\fn_{q}^{2}$
in the second row of $\mathbf{K}_{p q}^{M}$. We have shown by
explicit calculation that the presence of these extra
$1/\fn_{q}^{2}$ factors does not affect the calculation of spectral
singularities except for changing the value of the corresponding
mode numbers.

\end{document}